\documentstyle[preprint,aps]{revtex}
 \begin{document}
\preprint{USACH/99/01 , DFTUZ/99-02}
\title{Light-cone quantization of two dimensional\\
       field theory in the path integral approach}
\author{J.L. Cort\'es $^1$\thanks{E-mail: cortes@leo.unizar.es}
and J. Gamboa$^2$\thanks{E-mail: jgamboa@lauca.usach.cl}, } 
\address{
$^1$Departamento de F\'{\i}sica Te\'orica, Universidad de
Zaragoza, Zaragoza 50009, Spain\\ 
$^2$Departamento de F\'{\i}sica, Universidad de Santiago de Chile, 
Casilla 307, Santiago 2, Chile }

\maketitle

\begin{abstract} 
A quantization condition due to the boundary conditions and the 
compatification of the light cone space-time coordinate $x^-$
is identified at the level of the classical equations for the
right-handed fermionic field in two dimensions. A detailed
analysis of the implications of the implementation of this 
quantization condition at the quantum level is presented.
In the case of the Thirring model one has selection rules 
on the excitations as a function of the coupling and in the 
case of the Schwinger model a double integer structure of the
vacuum is derived in the light-cone frame. Two different
quantized chiral Schwinger models are found, one of them without 
a $\theta$-vacuum structure.
A generalization of the quantization 
condition to theories with several fermionic fields and to
higher dimensions is presented.
\end{abstract}

\section{Introduction}

Light-cone quantization (LCQ) has been considered recently as an
appropriate framework for a non-perturbative study of relativistic
quantum field theory \cite{brodsky1,wilson,bigatti}. The structure 
of the vacuum and bound states are examples of ingredients which
have a very different description in LCQ compared with 
canonical equal-time quantization. This opens the possibility to
get a new perspective on basic problems like the mechanism of
confinement in non-abelian gauge theories by the use of different
quantization schemes. 

From a technical point of view a property of the quantization in 
the light cone which plays a crucial role in the present work is
that it is naturally defined over a manifold with non-trivial 
topology. This can easily be understood from the form of the 
dispersion relation in the light-cone $k_+k_- -k_\bot^2 =m^2$. 
In order to be able to get the light-cone energy $k_-$ from this 
expression one considers discrete light-cone quantization (DLCQ) 
\cite{brodsky2}, i.e., a compactified $x^-$ coordinate, 
$-L\leq x^- \leq L$, leading to discrete non-zero values for the 
light-cone longitudinal momentum $k_+$.  

The aim of this paper is to explore physical effects whose 
explanation is related to this implicit difference in topology  
from standard spacetime quantization. We concentrate the 
discussion on the simplest system, two-dimensional field theory,
and as a tool the path integral formulation, which has proved to
be very useful at least at a formal level in the derivation of 
the fundamental aspects of quantum field theory, is used. In the
next section we consider quantum
mechanics as a toy model to illustrate the 
general framework of the discussion. In section 3 we study the
Thirring model; following the standar representation of the
fermion self-coupling by an auxiliary vector field we find a
quantization condition on the integral of the vector field 
along a constant $x^-$ closed line. When the relation between
the auxiliary field and the fermion current is taken into account   
and one considers a coupling $g^2$ such that $g^2/\pi$ is a 
rational number then the quantization condition translates into
a restriction on the possible values of the number of left-handed 
fermions, a result obtained previously \cite{Topthi} by other 
means. In section 4 similar arguments are applied to the Schwinger
model and the same quantization condition of the Thirring model
is obtained with the auxiliary field replaced by the gauge field.
In this case there is no algebraic relation between the vector
field and the fermion density; as a consequence of the quantization
condition on the integral involving the $A_-$ component one finds
a restriction on the asymptotic behavior at $x^+$ equal to plus or
minus infinity for the zero mode ($x^-$ independent component) of
the scalar field whose derivative with respect to $x^+$ gives 
the $A_+$ component. This asymptotic behavior is given in terms of
a second integer leading to a double integer structure for the 
model. We end up in section 5 with a brief discussion of possible
generalizations including the case with more than one fermionic
field where the quantization condition appears as an equation
por the $x^-$ path ordered exponential of the integral of an
appropriate non-abelian $A_-$ field over the constant $x^-$ closed
line. 
   
\section{A quantum mechanical example}

In \cite{Topthi} we showed that for the Thirring model in the light cone, 
the composite $\int_{-L}^{L} dx^- \psi^{\dagger}_L \psi_L$ is quantized
and we argued also that this quantization is due to  $S^1\times \Re$
topology induced by the light cone frame. In this section we will show
that this quantization rule is quite natural and it is always present in
the quantization of physical systems  when periodic or antiperiodic
boundary conditions are assumed.

We illustrate our argument considering the quantum mechanics
of a particle with an action~\cite{Gil}
\begin{equation}
 S = \int_0^T dt \biggl[{{\dot x}^2\over 2} - 
V 
- \psi^\dagger (i \partial_t + W)\psi \biggr], \label{1}
\end{equation}
where $V=V(x),\,W=W(x)$ and $x$ and $\psi$ are bosonic and fermionic
degrees of freedom respectively. In order to compute 
$Z$ we must fix the boundary conditions as follows
\begin{eqnarray}
x(0) &=& x(T),         \label{pbc }\\
\psi(0) &=& - \psi(T), \label{apbc}
\end{eqnarray}
and the partition function after integration over fermionic degrees of
freedom becomes
\begin{equation}
  Z = \int {\cal D}x \,\, \det [ i\partial_t + W] \,\, 
  e^{- \int_{0}^{T} dt\,[ \frac{1}{2} {\dot x}^2 - V ]}.
\label{path}
\end{equation}

The fermionic determinant, as usual, is $\prod_n \lambda_n$ where
$\lambda_n$ are the eigenvalues of the one-dimensional Dirac operator
$i\partial_t + W$, namely 
\begin{equation}
  [ i\partial_t + W] \varphi_n = \lambda_n \varphi_n.
\label{eige}
\end{equation}				

Equation (\ref{eige}) can be formally integrated out and the exact
solution is 
\begin{equation}
\varphi (t) = \varphi (0)\, e^{ i \int_0^t dt^{'} ( W (x(t^{'})) -
\lambda_n)}, \label{auto}
\end{equation}
and using (\ref{apbc}), one finds 
\begin{equation}
 \int_0^{T} dt\,\, W - \lambda_n T = 2\pi (n + \frac{1}{2}),
\label{quant}
\end{equation}				
$\forall\, n \in\,\,{\bf Z}$.

On the other hand a quantization condition similar to  the flux
quantization condition found in \cite{Topthi} for the Thirring model can
be derived directly from the equation of motion  for the fermion variable
$\psi$
\begin{equation}
(i \partial_t + W)\psi = 0, \label{motion}
\end{equation}				
with the boundary condition $\psi (T) = - \psi (0)$; one finds the
condition 
\begin{equation}
\int_0^{T} dt \,\,W = 2\pi (m + \frac{1}{2}).  \label{quant1}
\end{equation}

However (\ref{quant}) and (\ref{quant1}) are compatible if and only if
$\lambda_n =0$ is one of the eigenvalues of the Dirac operator. 

The flux quantization condition gives a restriction on the possible
trayectories
$x(t)$. The path integral (\ref{path}) is decomposed in topological
sectors
and it is replaced by a sum (over $m$) of integrals of the trayectories 
satisfying (\ref{quant1}). In general one can introduce a new parameter 
$\theta$ and an additional weight $e^{im\theta}$ to each sector.

As a simple example that shows explicitly the physical meaning of the 
flux quantization condition, let us consider a linear function
$W=\lambda x$. In this case (\ref{quant1}) becomes

\begin{equation}
\int_{0}^{T} dt\,\, x(t) = \frac{2\pi}{\lambda}(m+\frac{1}{2})\,.\nonumber
\end{equation}
If one introduces the decomposition in modes

\begin{equation}
x(t) = x^{(0)} + \sum_{n\neq 0} a_n \sin (\frac{n\pi}{T} t)\,,\nonumber
\end{equation}
then the flux quantization condition fixes the zero mode $x^{(0)}$ in
terms
of the integer $m$

\begin{equation}
x^{(0)} = \frac{2\pi}{\lambda T} (m + \frac{1}{2})\,,\nonumber
\end{equation}
and the sum over topological sectors replaces the integral over the zero
mode
in the original path integral (\ref{path}). 

For other choices of $W$ one has similar results, a relation (which will
be more involved)  between  the zero mode $x^{(0)}$ and the integer $m$
but the physical meaning of the flux quantization condition (\ref{quant1})
is the same.

\section{Thirring model}
In \cite{Topthi} a first discussion of the possible implications of
a compactified light-like coordinate in the quantization of the 
Thirring model was presented. The study was based on the identification
of a relation of the integration over the auxiliary variable (in the 
ligth-cone quantization) $\psi_R$ with a quantum mechanical determinant.
Consistency of the determinant lead to a quantization of the product
of the left-handed charge $Q_L$ and the Thirring coupling $g^2$ and 
then to the identification of many more excitations than those 
expected perturbatively.

In this section a complementary analysis of the quantization of the
Thirring
model compatible with the conservation of the vector current is presented. 
The starting point is the lagrangian, 

\begin{equation}
{\cal L}^{Th}= \psi^{\dagger}_L (i\partial_+ +A_+)\psi_L +
\psi^{\dagger}_R (i\partial_- +A_-)\psi_R + {1\over 2g^2} A_+ A_-\,,
\label{LTh}
\end{equation} 
which, eliminating the  auxiliary vector field $A_\mu$, reproduces the
lagrangian of the massless Thirring model in ligth cone coordinates
$x^\pm = t \pm x$. In order to identify the consequences of the 
boundary conditions on the fermionic field at $x^- =\pm L$ we consider
the solution of the equation $(i\partial_- +A_-)\psi_R =0$ for the 
rigth-handed fermionic field

\begin{equation}
\psi_R (x^+,x^-) = e^{i\int_0^{x^-} dy^- A_-(x^+,y^-)} \psi_R (x^+,0)\,.
\label{psiR}
\end{equation}
The most general boundary conditions compatible with charge 
conjugation symmetry are either periodic boundary conditions
(p.b.c.) $\psi (x^+,L) = \psi (x^+,-L)$ or antiperiodic boundary
conditions (a.b.c) $\psi (x^+,L) = - \psi (x^+,-L)$. In order to have 
a solution for $\psi_R$ compatible with the boundary conditions, the 
auxiliary field component $A_-$ has to satisfy the restriction

\begin{eqnarray}
\int_{-L}^L dx^- A_{-}(x^+,x^-) &=& 2\pi N \,\,\,
\mbox{for (p.b.c.)}\,\,\,,\nonumber\\
\int_{-L}^L dx^- A_{-}(x^+,x^-) &=& 2\pi (N +{1\over 2}) 
\,\,\,\mbox{for (a.b.c.)}\,\,\,.\label{Fqc}
\end{eqnarray}
On the other hand if one eliminates the auxiliary field 
component $A_+$, one finds $A_- = -2g^2 \psi^{\dagger}_L \psi_L$
and the restriction on the auxiliary field component $A_-$
can be translated into a quantization condition on the left-handed
chiral charge $Q_L =\int_{-L}^L dx^- \psi^{\dagger}_L \psi_L$,

\begin{eqnarray}
Q_L &=& {\pi \over g^2} N \,\,\, \mbox{for (p.b.c.)}\,\,\,,\nonumber\\
Q_L &=& {\pi \over g^2} (N +{1\over 2}) \,\,\, \mbox{for (a.b.c.)}\,\,\,, 
\label{QLqc}
\end{eqnarray}
giving an important information on the non-perturbative spectrum of 
the model \cite{Topthi}.

One can also see how the standard analysis of the model based on
bosonization is modified due to the presence of a compact
light-like dimension. In this case one has to consider the 
effective fermionic action, $\Gamma_f [A]$, defined by

\begin{equation}
e^{i\Gamma_{f}[A_{+},A_{-}]}=\int d\psi^{\dagger}d\psi 
e^{i\int d^{2}x (\psi^{\dagger}_{L}(i\partial_{+}+A_{+})\psi_{L}+
\psi^{\dagger}_{R}(i\partial_{-}+A_{-})\psi_{R})}\,,
\label{FEA}
\end{equation}
together with the two-dimensional representation of the (auxiliary) 
vector field in terms of the derivatives of two scalar fields. Using
ligth-cone coordinates, with $-L\leq x^- \leq L$, one has 
$A_+ =\partial_+ \chi$, $A_- =\partial_- \phi$ with

\begin{eqnarray}
\chi(x^+,x^-) &=& \sum_n \chi_n (x^+) e^{{in\pi\over
L}x^-}\,\,,\nonumber\\
\phi(x^+,x^-) &=& \sum_{n\neq0} \phi_{n}(x^{+}) e^{{in\pi\over L}x^{-}}+
{A^{(0)}_{-}(x^{+}) \over 2L} x^{-}\,.
\label{sfme}
\end{eqnarray} 
We have made explicit in the expansion the zero mode $A^{(0)}_-$,

\begin{equation}
A^{(0)}_- = \int_{-L}^L dx^- A_-\,\,,
\label{zm}
\end{equation} 
which instead of being a function of $x^+$ becomes a discrete
variable as a consequence of the restriction (\ref{Fqc}) due
to the boundary conditions.

When the result for the fermionic effective action

\begin{equation}
\Gamma_{f}[A_{+},A_{-}]={1\over 2\pi}\int d^{2}x \left(
A_{+}A_{-}-{1\over 2}A_{+}{\partial_{-}\over \partial_{+}}A_{+}
-{1\over 2}A_{-}{\partial_{+}\over \partial_{-}}A_{-} \right)\,,
\label{FEA2}
\end{equation}
where the regularization arbitrariness has been fixed by conservation 
of the vector current, is combined with the $x^-$-expansion of $A_{\pm}$
one finds

\begin{equation}
\Gamma_{f} [\partial_{+}\chi ,\partial_{-}\phi] = 
 {A^{(0)}_{-}\over 2\pi} \Delta\chi_{0} + \sum_{n}{{i n}\over 4L}\int
dx^{+} \left(\chi_{-n}-\phi_{-n}\right)
\partial_{+}\left(\chi_{n}-\phi_{n}\right)\,\,,
\label{BFEA}
\end{equation}
where $\Delta \chi_{0}=\int dx^+ \partial_{+}\chi_{0}=\chi_{0}(\infty)-
\chi_{0}(-\infty)$. 

The bosonized action ${\cal S}^{Th}_{bos}$ is obtained by adding to 
(\ref{BFEA}) the term quadratic in the auxiliary field of the original 
lagrangian (\ref{LTh}),
\begin{eqnarray}
{\cal S}^{Th}_{bos}&=&\Gamma_{f}[\partial_{+}\chi ,\partial_{-}\phi]
+{1\over {2g^{2}}}\int d^{2}x \partial_{+}\chi\partial_{-}\phi \nonumber
\\ 
&=& \left(1+{\pi\over g^{2}}\right)\bigg[{A^{(0)}_{-}\over 2\pi}
\Delta\chi_{0} 
-\sum_{n}\left({{i n}\over 2L}\right)\int dx^{+}\phi_{-n}\partial_{+}
\chi_{n} \nonumber \\ 
&\times&\sum_{n}\left({{i n}\over 4L}\right)\int dx^+ 
\left(\chi_{-n}\partial_{+}\chi_{n}+
\phi_{-n}\partial_{+}\phi_{n}\right)\bigg].
\label{BThA}
\end{eqnarray} 
The bosonized action is  a sum of 
terms (non-zero modes) quadratic in the derivatives of two scalar fields
in a two dimensional space with $-L\leq x^- \leq L$ plus a contribution
of the zero modes which involves the discrete variable $A^{(0)}_{-}$ and
$\Delta\chi_{0}$ exclusively. All the implications of the compactification
of the light-like coordinate $x^-$ are contained in the contribution of
the zero modes. 

The first restriction on these zero modes comes from the relation

\begin{eqnarray}
\left({-g^{2}\over \pi}\right) n_{L} &=& N \,\,\,\mbox{for
(p.b.c.)}\,\,\,,
\nonumber\\
\left({-g^{2}\over \pi}\right) n_{L} &=& (N +{1\over 2}) \,\,\, 
\mbox{for (a.b.c.)}\,\,\,,
\end{eqnarray}
which is the quantization condition (\ref{QLqc}) with $Q_L$ replaced by 
the number $n_L$ of left-handed fermions at fixed $x^{+}$.

In the case of antiperiodic boundary conditions $g^{2}/\pi$ has to be 
a rational number with even denominator, $g^{2}/\pi=p/2q$ and the 
number of left-handed fermions has to be a multiple of $q$, 
$n_{L}= q {\hat n}_{L}$. The integer $N$ which fixes the value of the 
discrete variable $A^{(0)}_{-}$ (see (\ref{Fqc})) is given by 
\begin{equation}
N = -\left({{p {\hat n}_{L} + 1}\over 2}\right)\,.
\end{equation}
If one introduces an aditional weight factor $e^{i\theta N}$ to each 
sector, the sum over $N$ can be easily evaluated using the only term
which depends on the zero modes, 
$e^{i{A^{(0)}_{-}\over 2\pi} \Delta\chi_{0}}$; one finds a quantization 
condition on $\Delta\chi_{0}$

\begin{equation}
{p\over 2}\left[\left(1+{\pi\over g^{2}}\right)\Delta\chi_{0} 
+\theta \right] = 2\pi M \,,
\label{chiqc}
\end{equation}    
with integer M.

Taking into account that $A_{+}=-2g^2 \psi^{\dagger}_R \psi_R$
and the dependence of $\psi_{R}$ on $x^-$ given by the solution
(\ref{psiR}) for the right-handed field, one has an auxiliary 
field component $A_+$ which does not depend on $x^-$. This means 
that one can take $\chi_n =0$ in the mode expansion and also that
$\Delta\chi_{0}$ can be reexpressed in terms of the original 
fermionic variables:

\begin{equation}
\Delta\chi_{0}={1\over 2L}\int_{-L}^{L}dx^{-}\int dx^{+}A_{+}=
-2g^{2}\int dx^{+}\psi^{\dagger}_{R}\psi_{R}\,\,.
\end{equation} 
Then the quantization condition for $\Delta\chi_{0}$ (\ref{chiqc})
can be written in the form

\begin{equation}
{p\over 2} \left[{\theta\over 2\pi}-\left(1+{p\over 2q}\right) 
n_{R}\right] = M \,\,,
\end{equation}
where the integer $n_{R}$ is the 
number of right-handed fermions for fixed $x^-$. From this relation 
one concludes that $\theta/2\pi$ has to be also a rational number,
${\theta\over {2\pi}}={m\over {2pq}}$ and the number of 
right-handed fermions is restricted by the condition that
$m-p(p+2q)n_{R}$ has to be a multiple of $4q$.

In the case of periodic boundary conditions one has a trivial
solution to the quantization condition for the zero modes 
with $n_{L}=0$. In this case $N=0$ and there is no possibility
to introduce an angle $\theta$, $g^{2}/\pi$ can be any real
number and the zero modes do not appear in the action. 
Together with this trivial case there is another solution to
the quantization condition for the zero modes which requires
$g^{2}/\pi$ to be a rational number as in the case of
antiperiodic boundary conditions, $g^{2}/\pi = p/q$. In this 
case one also finds that the number of left-handed fermions is a 
multiple of $q$, $n_{L}=q {\hat n}_{L}$ and the integer 
$N$ is in this case $N = -p {\hat n}_{L}$. The quantization
condition for $\Delta\chi_{0}$ which results from the
sum over $N$ with an aditional weigth factor $e^{i\theta N}$
is

\begin{equation}
p \left[{\theta\over 2\pi}-\left(1+{p\over q}\right) 
n_{R}\right] = M \,.
\end{equation}
In this case one finds that $\theta/2\pi$ has to be a rational
number, ${\theta\over {2\pi}}={m\over pq}$ and the number of
right-handed fermions has to be such that $m - p^{2}n_{R}$
is a multiple of $q$.

We end up this section emphasizing that the light-cone quantization
provides new non-perturbative information that in the usual spacetime
quantization we cannot see. In particular, the existence of new
excitations, the zero modes and the preponderant role played  by the
boundary conditions are new ingredients that are absent in the spacetime
quantization  
\section{Schwinger model}
The analysis in the previous section of the implications on the
quantization  
of a compactified light-like coordinate for the Thirring model can be 
easily applied to the Schwinger model. The discussion of boundary
conditions based on the solution (\ref{psiR}) for $\psi_R$ and the
expansion of the fermionic effective action (\ref{BFEA}) can be 
directly applied with the replacement of the auxiliary vector field of the
Thirring model by the dynamical gauge field in the Schwinger model.
Then, instead of the quadratic term in the auxiliary vector field of
the Thirring model, one has the action of the gauge field

\begin{equation}
{\cal S}_{g}={1\over {4e^{2}}}\int d^{2}x 
\left(\partial_{+}A_{-}-\partial_{-}A_{+}\right)^{2}\,\,.
\label{gfa}
\end{equation}
Using the $x^{-}$-expansion of the gauge field one has 

\begin{equation}
\partial_{+}A_{-}-\partial_{-}A_{+}=-\sum_{n}\left({in\pi\over L}\right)
\partial_{+}(\chi_{n}-\phi_{n})e^{{in\pi\over L}x^{-}}\,\,.
\end{equation}
The sum of the fermionic effective action (\ref{BFEA}) and the gauge 
field action (\ref{gfa}) leads to the bosonized action of the Schwinger 
model

\begin{eqnarray}
{\cal S}^{SM}_{bos}&=& {A^{(0)}_{-}\over 2\pi} \Delta\chi_{0} +  
{1\over 4\pi}\sum_{n}\left({{in\pi}\over L}\right)\int dx^+ 
\left(\chi_{-n}-\phi_{-n}\right)\partial_{+}\left(\chi_{n}-\phi_{n}\right)
-\nonumber\\
&&{1\over {4e^{2}}}\sum_{n}\left({{in\pi}\over L}\right)^{2}\int dx^+
\partial_{+}\left(\chi_{-n}-\phi_{-n}\right)
\partial_{+}\left(\chi_{n}-\phi_{n}\right)\,.
\label{BSchA}
\end{eqnarray} 

Together with the zero mode contribution already found in the 
Thirring model, we have an action for the combination $\chi-\phi$
of the two scalar fields which appear in the two-dimensional
decomposition of the gauge field. The fact that only one combination 
of the two scalars fields appear in the action is a consequence
of gauge invariance. From the relative coefficient of the two terms
involving $\chi_{n}-\phi_{n}$ one reproduces the well known mass gap
of the model ($m^{2}=e^{2}/\pi$).

It is also clear that introducing an external source $J_{\mu}$ one has

\begin{equation}
\partial_{\mu} <{\bar \psi}\gamma^{\mu}\psi> =
\left[\partial_{-} i {\delta\over {\delta J_{-}}} +
\partial_{+} i {\delta\over {\delta J_{+}}}\right] 
e^{i\Gamma_{f}[A+J]}\vert_{J=0} \,\,=\,\,0
\end{equation}
and
\begin{equation}
\partial_{\mu} <{\bar \psi}\gamma^{\mu}i\gamma_{5}\psi> =
\left[\partial_{-} i {\delta\over {\delta J_{-}}} -
\partial_{+} i {\delta\over {\delta J_{+}}}\right] 
e^{i\Gamma_{f}[A+J]}\vert_{J=0} \,\,=\,\,{e\over \pi}
(\partial_{+}A_{-}-\partial_{-}A_{+})\,,
\end{equation}
which reproduces the conservation of the vector current and the 
anomaly in the axial current.

In the Schwinger model there is no analog of the relation found in the
Thirring model between the (auxiliary) vector field and the fermionic
variables. Then the discussion of the zero modes is very simple. There
is no restriction on the integer $N$ which appears in the quantization
condition (\ref{Fqc}) on the zero mode of the gauge field component 
$A_{-}$. On the other hand if one sums over all posible values of $N$
with an additional relative weight $e^{i\theta N}$ then one has a 
simple quantization condition for $\Delta \chi_{0}$

\begin{equation}
\Delta\chi_{0}+\theta=2\pi M\,\,.
\label{A+qc}
\end{equation}
It is remarkable that, although one has 

\begin{equation}
\int d^{2}x F_{01} = -\sum_{n}\left({in\pi\over L}\right)
\int dx^{+} \partial_{+} \left(\chi_{n}-\phi_{n}\right)
\int dx^{-} e^{{in\pi\over L}x^{-}} = 0\,,
\end{equation}
and then apparently a trivial vacuum structure, the two integer 
structure of the Schwinger model \cite{Abdalla} is easily derived 
from the contribution of the zero modes ($N$ corresponding to the
zero mode of the $A_-$ component and $M$ corresponding to the 
zero mode of $A_{+}$).

The analysis of the implications of a light-like compact coordinate
in the quantization of a theory with a dynamical vector field can
also be applied to the chiral Schwinger model. There are two
different models 

\begin{eqnarray}
{\cal L}^{CSM}_{L} &=& \psi^{\dagger}_{L}\left(i\partial_{+}-A_{+}\right)
\psi_{L} + \psi^{\dagger}_{R}i\partial_{-}\psi_{R} 
- {1\over 4} F_{\mu\nu}F^{\mu\nu}\,,\nonumber\\
{\cal L}^{CSM}_{R} &=& \psi^{\dagger}_{L}i\partial_{+}\psi_{L}
+ \psi^{\dagger}_{R}\left(i\partial_{-}-A_{-}\right)\psi_{R} 
 - {1\over 4} F_{\mu\nu}F^{\mu\nu}\,.
\end{eqnarray}
In the first case the equation for the right-handed field does not
involve the vector field and there is no quantization condition
on the zero modes of the vector field. The effective fermionic 
action is given in this case by 

\begin{equation}
\Gamma^{L}_{f}[A_{+},A_{-}]={1\over 2\pi}\int d^{2}x \left(
a A_{+}A_{-}-{1\over 2}A_{+}{\partial_{-}\over \partial_{+}}A_{+}
\right)\,,
\label{LFEA}
\end{equation} 
where, as a consequence of the chiral coupling, the non-local 
contribution involves only the $A_{+}$ component and the local term 
has a coeficient with a regularization dependent parameter $a$ due 
to the lack of gauge invariance \cite{Jackiw}. The bosonized 
action, obtained by using the decomposition of the vector field
in terms of two scalar fields (\ref{sfme}), is

\begin{eqnarray}
&&{\cal S}^{LCSM}_{bos}={a\over 2\pi}\int dx^{+}A^{(0)}_{-}\chi_{0} -
{a\over 2\pi}\sum_{n}\big({in\pi\over L}\big)\int dx^{+}
\phi_{-n}\partial_{+}\chi_{n}
+ \nonumber\\ {1\over 4\pi}\sum_{n}\big({in\pi\over L}\big)&&\int dx^{+}
\chi_{-n}\partial_{+}\chi_{n} -
{1\over {4e^{2}}}\sum_{n}\left({{in\pi}\over L}\right)^{2}\int dx^+
\partial_{+}\left(\chi_{-n}-\phi_{-n}\right)
\partial_{+}\left(\chi_{n}-\phi_{n}\right)\,.
\end{eqnarray}   
Note that in this case there is no quantization condition on the zero mode
of $A_-$ and $\theta$-vacua are absent, a result derived
recently~\cite{Prem}
in the light-cone hamiltonian formulation.
 
In the chiral model with an interacting right-handed component one 
has the same quantization condition (\ref{Fqc}) on the zero mode of 
$A_{-}$ as in the Schwinger model and the bosonized action is

\begin{eqnarray}
&&{\cal S}^{RCSM}_{bos}={a\over 2\pi} A^{(0)}_{-}\Delta\chi_{0} -
{a\over 2\pi}\sum_{n}\big({in\pi\over L}\big)\int dx^{+}
\phi_{-n}\partial_{+}\chi_{n}
+ \nonumber\\{1\over 4\pi}\sum_{n}\big({in\pi\over L}\big)&&\int dx^{+}
\phi_{-n}\partial_{+}\phi_{n} -
{1\over {4e^{2}}}\sum_{n}\left({{in\pi}\over L}\right)^{2}\int dx^+
\partial_{+}\left(\chi_{-n}-\phi_{-n}\right)
\partial_{+}\left(\chi_{n}-\phi_{n}\right)\,.
\end{eqnarray} 
The treatment of the zero modes is similar to the gauge invariant case 
with the quantization condition for $\Delta\chi_{0}$ (\ref{A+qc})
replaced by

\begin{equation}
a\Delta\chi_{0}+\theta=2\pi M\,.
\end{equation}

\section{Discussion}
We end up by considering the possibility to extend the analysis of
the Thirring model, Schwinger model and chiral Schwinger models presented
in this work. One property of all these models crucial in all the 
discussion is the absence of a mass term for the fermion allowing to
decouple the two chiral components in the equation for the fermion
field. With this restriction there are two possible extensions to
consider. 

The first generalization is to consider a non-abelian
generalization where one has several fermionic fields instead of
a single Dirac field. One will have an equation for the right-
handed fermion field multiplet with the vector field $A$ in the 
abelian cases replaced by a Lie algebra valued vector field and 
also a generalization of the boundary condition for the 
rigth-handed field $\psi_{R}(x^{+},L)=e^{2\pi i\Gamma} \psi_{R}(x^{+},-L)$
with a matrix $\Gamma$ in internal space. One will have in this case
a non-abelian generalization of the quantization condition (\ref{Fqc}),

\begin{equation}
{\cal P}_{x^{-}} e^{i\int_{-L}^{L} dx^{-} A_{-}} = e^{2\pi i\Gamma}\,\,,
\end{equation}
where ${\cal P}_{x^{-}}$ denotes a path ordering on the light-like
coordinate $x^{-}$.
The study of the consequences of this quantization condition in a
non-abelian generalization of the Thirring model as well as in
models with dynamical vector fields deserves further investigation.

A second generalization of the present work is to try to go
beyond two space-time dimensions. In higher dimensions the solution
of the right-handed fermion field equation can be expressed also
as a path-ordered exponential

\begin{equation}
\psi_{R}(x^{+},x^{-},{\vec x}_{\perp}) =
{\cal P}_{x^{-}}e^{i\int_{0}^{x^{-}}dy^{-}
\big[A_{-}(x^{+},y^{-},{\vec x}_{\perp})-i{\vec \alpha}_{\perp}
\big({\vec \partial}_{\perp}+
i{\vec A}_{\perp}(x^{+},y^{-},{\vec x}_{\perp})\big)\big]}
\psi_{R}(x^{+},0,{\vec x}_{\perp})\,. 
\end{equation}
Then the choice of periodic or antiperiodic boundary conditions for
the right-handed fermionic field in light-like coordinate $x^{-}$
leads to a higher dimensional quantization condition

\begin{equation}
{\cal P}_{x^{-}}e^{i\int_{-L}^{L}dy^{-}
\big[A_{-}(x^{+},y^{-},{\vec x}_{\perp})-i{\vec \alpha}_{\perp}
\big({\vec \partial}_{\perp}+
i{\vec A}_{\perp}(x^{+},y^{-},{\vec x}_{\perp})\big)\big]}
= \pm 1\,\,,
\end{equation}
which is more difficult to analyze than the two dimensional
case due to the presence of the additional transverse 
coordinates ${\vec x}_{\perp}$.
This should be the starting point for a study of the consequences 
of a compactified light-like coordinate $x^{-}$ on the quantization 
of a theory with a massless fermionic field.       

\bigskip

{\bf Acknowledgements}

\medskip
We would like to thank J. Zanelli by several interesting discussions about
this paper.This work was partially supported by CICYT (Spain) project
AEN-97-1680 and grants 1980788, 7980045 from FONDECYT-Chile, and
DICYT-USACH.


\begin{references} 
\bibitem{brodsky1} S. Brodsky, H.-C. Pauli and S. Pinsky, hep-ph/9705477, 
{\it Phys. Rep.} {\bf 301}, 299 (1998) and references therein.
\bibitem{wilson} M. M. Brisudova, R. Perry and K.G. Wilson,
hep-ph/9607280, 
{\it Phys. Rev. Lett.} {\bf 78}, 1227 (1997); S. D. Glazek and K. G.
Wilson, 
hep-th/9707028,
{\it Phys. Rev.} {\bf D57}, 3558 (1998).
\bibitem{bigatti} D. Bigatti and L. Susskind, hep-th/9711063, {\it Phys.
Lett. }{\bf B425}, 351 (1998).
\bibitem{brodsky2} T. Maskawa  and  K. Yamawaki, {\it Prog. Theor. Phys.}
{\bf 56} 270 (1976);
S. Brodsky and H. -C. Pauli, {\it Phys. Rev.}{\bf 32}, 
1993 (1985).
\bibitem{Topthi} J.L. Cort\'es, J. Gamboa, I. Schmidt and J. Zanelli,
hep-th/9809051, {\it Phys. Lett. }{\bf B444}, 451 (1998).
\bibitem{Gil} E. Gildener and A. Patrascioiu,
{\it Phys. Rev.} {\bf D16}, 1802 (1977).
\bibitem{Abdalla} E. Abdalla, M.B. Abdalla and K.D. Rothe,
{\sl Non-perturbative methods in 2 dimensional quantum field theory} pg.
301,
(World Scientific, Singapore, 1991)  
\bibitem{Jackiw} 
J.~Jackiw and R.~Rajamaran, {\it Phys. Rev. Lett.} {\bf 54}, 1219 (1985).
\bibitem{Prem}
P.P. Srivastava, hep-th/9811225, {\it Phys. Lett.}{\bf B} (in press).
\end{references}
\end{document}